\documentstyle[a4,12pt,epsf]{article}
\epsfverbosetrue
\newcommand{\figurebox}[2]{\fbox{\vbox to#2in{\hbox to #1in{\hfil} \vfil}}}
\newcommand{\ie} {{\em i.e.}}

\newcommand{\beq}{\begin{equation}}
\newcommand{\eeq}{\end{equation}}
\newcommand{\dslash}{\! \not \!\! D}

\newcommand{\ewxy}[2]{\setlength{\epsfxsize}{#2}\epsfbox[10 30 640 590]{#1}}
\newcommand{\ehxy}[2]{\setlength{\epsfysize}{#2}\epsfbox[0 30 690 590]{#1}}
\begin{document}

\begin{titlepage}

\begin{flushright}
Edinburgh Preprint: 92/509\\
\today
\end{flushright}
\vspace{5mm}
\begin{center}
{\Huge Hopping Parameter Expansion for Heavy-Light Systems}\\[15mm]

{\large\it UKQCD Collaboration}\\[3mm]
{\bf D.S.~Henty and R.D.~Kenway}\\
Department of Physics, The University of Edinburgh, Edinburgh EH9~3JZ,
Scotland
\end{center}
\vspace{5mm}
\begin{abstract}

We present a technique which permits the calculation of
two-point functions of operators containing one heavy quark and an arbitrary
number of light quarks as analytic
functions of the heavy-quark mass. It is based on the standard
Jacobi linear solver used for the calculation of quark
propagators. Results for the heavy-light pseudoscalar and vector meson
masses are obtained on $16^3 \times 48$ lattices at $\beta$ = 6.2 using the
Wilson fermion action, and agree
with published data. The incorporation of smeared operators and $O(a)$-
improved actions presents no problems.

\end{abstract}

\end{titlepage}

\paragraph{Introduction}\label{introduction}

There is much interest at present in studying systems involving both heavy
and light quarks on
the lattice. The reasons for this are twofold. Firstly,
the values of quantities such as the decay constants of the $D$
and $B$ mesons and the form of the Isgur-Wise function \cite{isgur-wise}
are currently of great interest to both
the experimental and theoretical communities. These are,
in principle, calculable on the lattice.
Secondly, it is expected that discretisation errors are relatively large
for states containing heavy quarks and such systems provide an
arena in which to look for gains from the use of
improved actions \cite{improvement}.

Although nature has so far only revealed two heavy quarks, charm and
bottom, with masses of approximately 1.5 and 5 GeV respectively,
it is not always
sufficient to perform lattice simulations solely at these two points.
Regarding the evaluation of the Isgur-Wise function $\xi({\em v}.{\em v}')$,
it is essential to know when (and if)
the heavy-quark effective theory is valid. Once this has been determined,
it is necessary to perform measurements at many different heavy-quark
velocities to map out the form of $\xi({\em v}.{\em v}')$.
Since the lattice quantises the allowed values of momenta, this can only be
done
in a continuous fashion by varying the quark mass.
In order to study lattice artifacts we need to have results at many
heavy-quark masses to discover when and how the discretisation errors manifest
themselves and to decide when to trust the lattice results.

In this letter, we use the standard Wilson fermion action
\begin{eqnarray}
\label{eq:wilson_action}
\nonumber
S_F^W & = & \sum _x \left\{\bar{q}(x)q(x)
            -\kappa\sum _\mu\left[
            \bar{q}(x)(1 - \gamma _\mu)U_\mu (x) q(x+\hat\mu ) +
            \bar{q}(x+\hat\mu )(1 + \gamma _\mu )U^\dagger _\mu(x)
            q(x)\right] \right\} \\
      & = & \bar{q} \left( 1 - \kappa \dslash \right) q
\end{eqnarray}
where $\kappa$, the quark hopping parameter, is
inversely related to the bare mass.
We develop a method which allows the hadron correlators to
be evaluated as convergent power series in the heavy-quark hopping
parameter. In what follows we denote generic quarks and
their hopping parameters by $q$, $\bar{q}$ and $\kappa$, and use $h$, $\bar{h}$
and $\kappa_h$ for heavy quarks.

\paragraph{Jacobi Algorithm}\label{jacobi}

To evaluate any lattice quark propagator $G$ we need to solve the
linear equation
\begin{equation}
\label{eq:prop_eqn}
(1 - \kappa \dslash ) G = \eta
\end{equation}
for $G$ with some given source $\eta$. The Jacobi algorithm comes from
the na\"{\i}ve Taylor expansion of the inverse of the fermion matrix, \ie
\begin{equation}
\label{eq:taylor_series}
G = \eta + \sum_{n=1}^{\infty} \kappa^n \dslash^n \eta .
\end{equation}
In practice, it is implemented by the recursion relation
\begin{eqnarray}
\label{eq:jacobi_alg}
\nonumber G_0 & = & \eta \\
G_{n+1} & = & \eta + \kappa \dslash G_n
\end{eqnarray}
where we terminate the series at some finite $n_{\rm max}$. The minimum value
of
$n_{\rm max}$ needed to achieve some desired accuracy can be chosen, for
example,
by monitoring the norm of the residue vector
\begin{equation}
\label{eq:residue}
{\em r}_n = \eta - (1 - \kappa \dslash) G_n .
\end{equation}
The algorithm is convergent for sufficiently small $\kappa$, and for the
heavy-quark masses of interest there is no problem with convergence.
For our purposes it is sufficient to notice that at iteration $n$, $G_n$
is updated by a term of order $\kappa^n$.

Dividing the lattice into even and odd sites and using the fact that $\dslash$
only connects even sites to odd sites
it is apparent that the solution $G$ of equation~(\ref{eq:prop_eqn})
satisfies
\begin{equation}
\label{eq:recover_odd_solution}
G_{\rm odd} = \eta_{\rm odd} + \kappa \dslash G_{\rm even}
\end{equation}
We need only compute the solution on even sites which satisfies
\begin{equation}
\label{eq:rb_prop_eqn}
(1 - \kappa^2 \dslash^2 ) G_{\rm even} = \eta_{\rm even} + \kappa \dslash
\eta_{\rm odd}
\end{equation}
Equations~(\ref{eq:recover_odd_solution}) and~(\ref{eq:rb_prop_eqn})
comprise the red-black preconditioning of equation~(\ref{eq:prop_eqn})
and for the Jacobi algorithm reduce the computational load by a factor of two.

\paragraph{Hopping Parameter Expansion}\label{hopping}

For simplicity, consider a mesonic operator
${\cal O}_{\Gamma} = \bar{h}\Gamma q$.
The quantity of interest is the timesliced correlator
$C_{\Gamma}(t)$ defined by
\begin{eqnarray}
\label{eq:tslice_corr_prop}
\nonumber
C_{\Gamma}(t) & = & \sum_{\vec x} \langle {\cal O}_{\Gamma}({\vec x}, t)~{\cal
O}^{\dag}_{\Gamma}({\vec 0}, 0) \rangle \\
              & = & \sum_{\vec x} \langle {\rm Tr} \left\{ \gamma_5\Gamma
G_q({\vec x}, t) \bar\Gamma \gamma_5 G_h^{\dag}({\vec x}, t)\right\} \rangle
\end{eqnarray}
where $\bar\Gamma = \gamma_0 \Gamma^{\dag} \gamma_0$.
We assume that the computationally intensive part of this calculation (the
evaluation of the light quark propagator $G_q$) has been performed.
Using the Jacobi algorithm, the standard approach would be to solve the
red-black preconditioned equation~(\ref{eq:rb_prop_eqn})
for some
particular value of $\kappa_h$ and compute $G_h$. $C_{\Gamma}(t)$ would
then be evaluated and the whole procedure repeated for each different value
of $\kappa_h$.

We introduce the hopping parameter expansion \cite{hasenfratz} by
inserting the Taylor series expansion for $G_h$
directly into equation~(\ref{eq:tslice_corr_prop})
to obtain $C_{\Gamma}(t)$ as a power series in $\kappa_h$:
\begin{equation}
\label{eq:corr_expansion}
C_{\Gamma}(t) \simeq \sum_{n=0}^{n_{\rm max}} c_{\Gamma}^{(n)}(t) \kappa_h^n
\end{equation}
where
\begin{equation}
\label{eq:coeff_def}
c_{\Gamma}^{(n)}(t) = \sum_{x} \langle {\rm Tr} \left\{ \gamma_5\Gamma G_q(x,
t) \bar \Gamma \gamma_5 \left[ \dslash^{n} \eta \right]^{\dag}(x, t)\right\}
\rangle
\end{equation}
and $\dslash^0 \eta = \eta$.

The computational overhead is the calculation of the
desired traces at every order of the expansion
rather than only once for the sum of the series.
The gain is that the light quark propagator only has to be read in to memory
once, which may be significant on certain machines for large lattices,
and results are obtained for a continuous range of $\kappa_h$.

It is obvious that the static approximation (the limit
of infinite quark mass / zero $\kappa_h$) is a simple by-product
of this more general finite mass expansion. The relationship is
\begin{equation}
C_{\Gamma}^{\rm static}(t) = c_{\Gamma}^{(t)}(t).
\end{equation}

The numerical problem with equation~(\ref{eq:corr_expansion})
as it stands is that the coefficients
$c_{\Gamma}^{(n)}(t)$ diverge as $n$ becomes large and eventually cause
an overflow on the computer. We can predict the growth of
$c_{\Gamma}^{(n)}(t)$ by the following simple argument.

Consider the pseudoscalar channel ($\Gamma = \gamma_5$). At
some critical value of the hopping parameter $\kappa = \kappa_c$, the
pseudoscalar comprising two degenerate light quarks (the pion) becomes
massless.
In this case $C_{\Gamma}(t)$ no longer decays exponentially
and so we expect all orders of the hopping parameter expansion to
contribute with roughly equal weight, \ie
\begin{equation}
c_{\gamma_5}^{(n)}(t) \kappa_c^n \sim {\rm constant}.
\end{equation}
Since the pseudoscalar is the lightest channel,
$c_{\gamma_{5}}^{(n)}(t)$ will have the worst divergences. If we can keep these
under control, then all other channels will be numerically stable as well.
Thus, although in practice we deal with light quark propagators which
correspond more to the strange quark mass than the chiral limit, if
we rewrite equation~(\ref{eq:corr_expansion}) as
\begin{equation}
\label{eq:corr_alpha_expansion}
C_{\Gamma}(t) \simeq \sum_{n=0}^{n_{\rm max}} \bar{c}_{\Gamma}^{(n)}(t)
\left[\alpha \kappa_h\right]^n
\end{equation}
with $\alpha \sim \kappa_c^{-1}$ then we expect the new expansion coefficients
$\bar{c}_{\Gamma}^{(n)}(t)$ to be roughly constant.

\begin{figure}
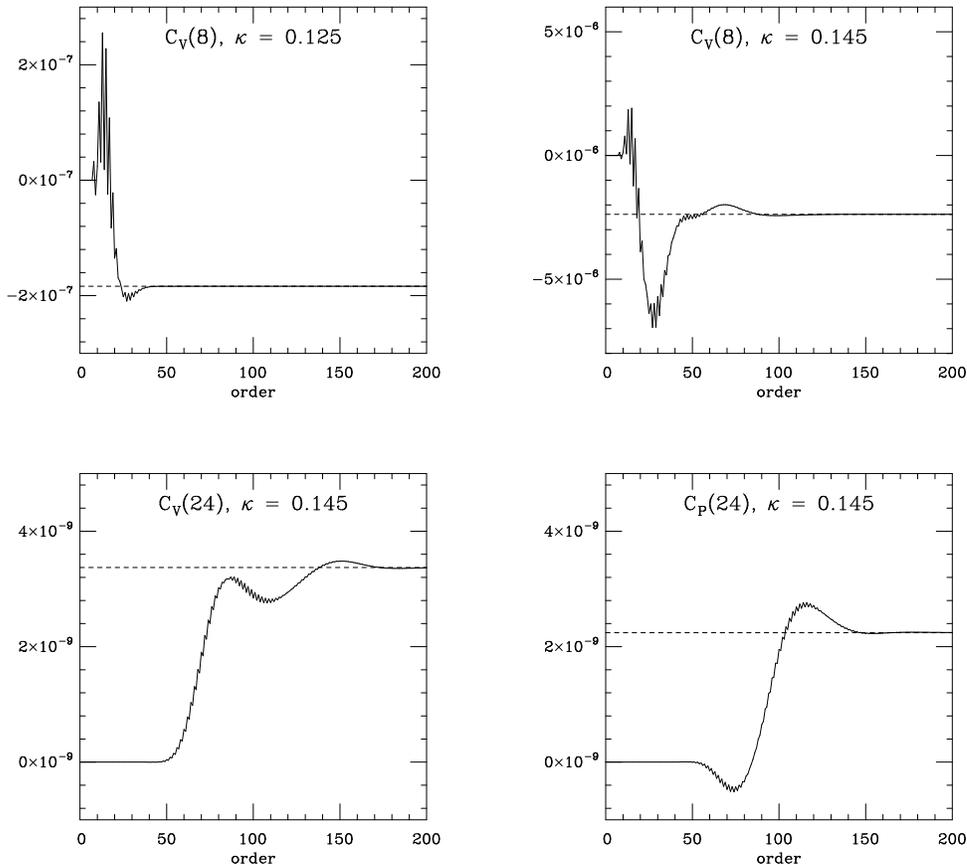

\centerline{	\ewxy{agreement1.ps}{175pt} \hspace{20pt}
		\ewxy{agreement2.ps}{175pt}}
\vspace{20pt}
\centerline{	\ewxy{agreement3.ps}{175pt} \hspace{20pt}
		\ewxy{agreement4.ps}{175pt}}
\caption{Convergence of Vector ($C_V$) and Pseudoscalar ($C_P$)
Meson Correlators\label{fig:agreement}}
\end{figure}

\paragraph{Simulation Details}
Numerical studies were performed on 10 lattices of size $16^3 \times 48$,
generated in the quenched approximation at $\beta$ = 6.2.
Since the major purpose of this letter is
to demonstrate the validity of the technique and to reproduce known results,
only two meson operators were calculated:
\begin{eqnarray}
\nonumber {\rm Pseudoscalar:} & \Gamma = & \gamma_5 \\
\nonumber {\rm Vector:}       & \Gamma = & \sum_{i=1}^3 \gamma_i
\end{eqnarray}
from which the masses and hyperfine splitting can be calculated and compared
to reference \cite{southampton}. The light-quark hopping parameter was chosen
as
$\kappa$ = 0.1510, corresponding roughly to the strange quark mass
\cite{UKQCD}, and these light propagators were
calculated using an over-relaxed minimal residual algorithm.
To span the range of heavy-quark masses in \cite{southampton}
($0.125 \leq \kappa_h \leq 0.145$),
the hopping parameter expansion was computed to $200^{\rm th}$ order.
This number was chosen
by explicitly calculating quark propagators at $\kappa_h = 0.145$ using the
Jacobi algorithm and requiring that the residue vector of
equation~(\ref{eq:residue}) satisfy $|r_{n_{\rm max}}| < 10^{-7}$.
The following test was performed on a single configuration.
Firstly, the standard propagator code (used to evaluate the light propagator)
was used to compute
the heavy-quark propagators at $\kappa_h = 0.145$ and $\kappa_h = 0.125$,
and the values of $C_\Gamma(t)$ calculated directly.
Secondly, the coefficients $\bar{c}_{\Gamma}^{(n)}(t)$ were computed
and the series summed off-line at the same two values
of $\kappa_h$. The results are illustrated in figure~\ref{fig:agreement},
where the solid line is the hopping parameter expansion summed
to a given order and the dashed line is the value of $C_\Gamma(t)$.
The results of these two independent calculations agreed, and it
was deduced that the hopping parameter expansion had converged
(to the accuracy desired) within the required range. As expected, convergence
is most rapid at small values of $\kappa_h$ and $t$. The other notable
features are that the convergence of neither
the vector nor the pseudoscalar meson correlators is monotonic, and that
the full 200 orders are necessary at the lightest mass and furthest timeslice.
For heavy quarks near the charm mass ($\kappa_h \simeq 0.135$ \cite{charm})
around 100 orders are sufficient.

The value of $\alpha$ was varied, and the scaled coefficients
$\bar{c}_{\Gamma}^{(n)}(t)$ found to be numerically stable for $\alpha = 6.0$.
The bare coefficients $c_{\gamma_5}^{(n)}(t)$ actually grow like $(6.3)^n$
which,
since the critical hopping parameter $\kappa_c \simeq 0.1533$ \cite{UKQCD},
agrees with our prediction $c_{\gamma_5}^{(n)}(t) \sim (6.5)^n$.

For these initial studies, point sources at the origin (an even site) and
point sinks were used for both the heavy and the light quarks.
In this case, the hopping parameter expansion is an implementation of the
red-black preconditioned Jacobi algorithm.
Although our tests, performed on a single configuration, were sufficient to
prove that the algorithm works we decided to use smeared operators
during the full simulation in order to obtain better signals for the relevant
particles.
Point sources were still used for the heavy quarks so that the red-black
factorisation could be exploited. The light-quark operators were smeared
at the source and the sink using a gauge covariant smearing function
with an RMS smearing radius of about four lattice spacings \cite{smearing}.
The incorporation of non-local heavy-quark sources and sinks is discussed
later.

\paragraph{Results}

Since in principle the simulation provides $C_\Gamma(t)$ as an analytic
function
of $\kappa_h$ there are several analysis methods available. Previously,
when the expansion was truncated well before convergence, the behaviour
of the Fourier transformed correlation functions was studied and the
particle masses identified as poles in momentum space \cite{hasenfratz}.
Our approach is to sum the series at many values of $\kappa_h$ and
fit the asymptotic exponential decay in time.

After the use of smeared operators, plateau regions in $C_\Gamma(t)$
were readily identified for the whole range of $\kappa_h$,
although the limited statistics meant that the signal deteriorated rapidly
especially at very large heavy-quark masses (here we approach the static quark
limit where this effect is well known \cite{static}).
An uncorrelated fit to the vector and pseudoscalar meson channels was done
from time slices 5 to 10 and the corresponding masses extracted.
To estimate the errors, the whole procedure was bootstrapped and the
probability distributions of the masses and mass difference obtained.
The quoted errors are 68\%
confidence level regions and are therefore not necessarily symmetric.
Unfortunately, lack of statistics prevented a proper correlated fit.

\begin{figure}
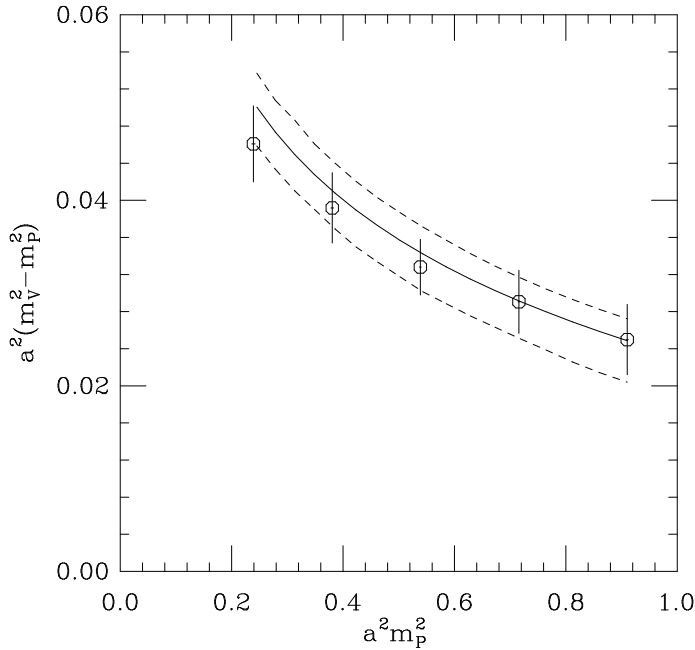

\centerline{\ehxy{splitting.ps}{250pt}}
\caption{Vector-Pseudoscalar Hyperfine Splitting \label{fig:hyperfine}}
\end{figure}

The most interesting quantity is $m_V^2 - m_P^2$, which becomes constant
in the heavy-quark effective theory with corrections of order $(m_V +
m_P)^{-1}$.
Experimentally, this quantity is remarkably constant all the way
from the $\pi, \rho$ system through the $K$ and $D$ to the $B$, with
a measured value of around 0.55 ${\rm GeV}^2$. We plot this mass difference
against $m_P^2$ in figure~\ref{fig:hyperfine} together with the data from
\cite{southampton}. Our analysis was actually done
at 17 evenly spaced values of $\kappa_h$, and we have drawn a continuous line
representing the best fits and dashed lines representing the envelope of
the error bars.
It is clear that, within errors, there is no discrepancy
between the two data sets.
The results show the well known feature that the hyperfine splitting for
heavy-quark mesons is far too small for the Wilson action (at $\beta = 6.2$,
$a^{-1}$ = 2.73 GeV \cite{UKQCD}) and decreases with
increasing heavy-quark mass, contrary to experiment. This is
observed both in heavy-light and heavy-heavy systems.

\paragraph{Non-Local Operators and Improved Actions}
It is not practical to smear the heavy-quark propagator at the sink since
this would have to be done at each order of the hopping parameter expansion,
taking a prohibitively long time for even the simplest of smearing techniques.
If smearing is required at the sink, it should be performed on the light
quark fields as this only needs to be done once.
The only complication arises if smearing is required at the source because the
only light-quark propagators available originate from point sources. In this
case, it is necessary to smear the heavy quark at the source. There is no
problem in doing this, the only disadvantage being that the red-black
factorisation cannot be exploited.

It has been suggested that the unphysically small value obtained for the
hyperfine splitting in heavy-quark systems is due to large order $a$ effects
present in the Wilson action. Using an $O(a)$-improved fermion action would
be expected to improve the situation, and this has been demonstrated for
the ``clover'' action by studying the charmonium system \cite{aida,charm}.
The important point is that $\kappa$ still only enters the fermion action
as an overall prefactor so it is straightforward to use the hopping parameter
expansion. The other element of the clover action
is that the quark propagator needs to be rotated at source
and sink to complete the $O(a)$ improvement \cite{improvement}, so
there is no advantage to red-black factorisation.

\paragraph{Conclusions}

We have shown that the hopping parameter expansion is a viable method
for the study of heavy-light systems on the lattice. It provides results
as an analytic function of the heavy-quark mass provided that this mass
is larger than some lower bound. This lower bound can be reduced by going
to higher orders of the expansion, but for studies at $\beta = 6.2$ on a $16^3
\times 48$ lattice only 200 orders are needed for
$\kappa_h \le 0.145$. This is lighter than charm for which
$\kappa_h \simeq 0.135$ and where around 100 orders are sufficient.
The hopping parameter expansion, when used with an $O(a)$-improved action,
provides an attractive method for mapping out the behaviour
of meson decay constants and the Isgur-Wise function continuously at and above
the charm quark mass.
\paragraph{Acknowledgements}
This work was supported by the UK Science and Engineering Research Council
through the grants GR/G 32779 and GR/H 01069, and by the University of
Edinburgh and Meiko Ltd. We are particularly grateful to Mike Brown and Arthur
Trewe for arranging access to the University of
Edinburgh's Connection Machine 200
on which the bulk of the calculation was performed. We thank our colleagues
Ken Bowler, Brian Pendleton, David Richards, Chris Sachrajda, Jim Simone
and Alan Simpson for helpful discussions and Jim Simone for the use of his
hadron analysis program.


\begin{thebibliography}{99}

\bibitem{isgur-wise} N.~Isgur and B.~Wise, Phys.\ Lett.\ 237B (1990) 527.

\bibitem{improvement} B.~Sheikholeslami and R.~Wohlert, Nucl.\ Phys.\ B259
(1985) 572. \\
G.~Heatlie et al., Nucl.~Phys. B352 (1991) 266.

\bibitem{hasenfratz} P.~Hasenfratz and I.~Montvay, Phys.\ Rev.\ Lett.\ 50
(1983) 309. \\
P.~Hasenfratz and I.~Montvay, Nucl.\ Phys.\ B237 (1984) 237.

\bibitem{southampton} C.R.~Allton et al., Nucl.\ Phys.\ B372 (1992) 403.

\bibitem{UKQCD} UKQCD Collaboration, {\em Quenched Hadrons using Wilson and
$O(a)$-Improved Fermion Actions at $\beta=6.2$},
to be published in Phys.\ Lett.\ B.

\bibitem{charm} UKQCD collaboration, {\em Lattice Computations of the
$J/\psi - \eta_c$ Mass Splitting}, in preparation.

\bibitem{aida} A.X.~El-Khadra, {\em Charmonium with Improved Wilson Fermions
II: The Spectrum}, Fermilab preprint FERMILAB-CONF-92/10-T (1992).

\bibitem{smearing} UKQCD collaboration, {\em Gauge Invariant Smearing
of Wilson Fermions at $\beta = 6.2$}, in preparation.

\bibitem{static} G.~P.~Lepage, {\em Simulating Heavy Quarks},
invited talk at Lattice 91, Tsukuba, Japan, Nov 5-9, 1991

\end{thebibliography}
\end{document}